\documentclass[a4paper]{jps-cp}

\title{Pressure-Induced Restoration of the Reversed Crystal-Field Splitting in $\alpha$-Sr$_2$CrO$_4$}

\author{
Ryo \textsc{Takahashi}$^{1}$, 
Tomoki \textsc{Yamaguchi}$^{1}$, 
Koudai \textsc{Sugimoto}$^{2,3}$, 
Touru \textsc{Yamauchi}$^{4}$, 
Hiroya \textsc{Sakurai}$^{5}$, and 
Yukinori \textsc{Ohta}$^{1}$
}

\inst{
$^{1}$Department of Physics, Chiba University, Chiba 263-8522 , Japan \\
$^{2}$Center for Frontier Science, Chiba University, Chiba 263-8522, Japan \\
$^{3}$Department of Physics , Keio University, Yokohama 223-8522, Japan \\
$^{4}$Institute for Solid State Physics, University of Tokyo, Kashiwa 277-8581, Japan \\
$^{5}$National Institute for Materials Science, Tsukuba 305-0044, Japan
}

\email{ohta@faculty.chiba-u.jp}

\recdate{August 29, 2019}

\abst{
Motivated by an experimental finding that the successive phase transitions in 
$\alpha$-Sr$_2$CrO$_4$ observed at ambient pressure ceases to exist under high 
pressures, we carry out the density-functional-theory-based electronic structure 
calculations and demonstrate that the reversal of the crystal-field splitting reported 
previously is restored under high pressures, so that the orbital degrees of freedom 
disappears, resulting in the single phase transition that divides the system into 
high-temperature Mott insulating and low-temperature antiferromagnetic 
insulating phases.  
}

\kword{Sr$_2$CrO$_4$, crystal field splitting, orbital ordering, pressure effect}

\begin{document}

\maketitle

\section{Introduction}

The orbital degrees of freedom in transition-metal compounds have long been one of the major themes in the 
physics of strongly correlated electron systems \cite{khomskii}.  One of the recent examples is the origin of 
successive phase transitions observed in a layered perovskite $\alpha$-Sr$_2$CrO$_4$ \cite{sakurai}.  
This material is a Mott insulator, having the K$_2$NiF$_4$-type crystal structure with CrO$_6$ octahedra 
elongated along the $c$-axis of the lattice and with a $3d^2$ electron configuration \cite{kafalas,baikie}.  
Therefore, one would naturally expect that two electrons occupy the lowest doubly degenerate $t_{2g}$ 
orbitals forming an $S=1$ spin due to Hund's rule coupling, so that only the antiferromagnetic N\'eel ordering 
of $S=1$ spins occurs below the N\'eel temperature $T_\textrm{N}$, without any orbital ordering.  However, 
a recent experimental study \cite{sakurai} revealed that two phase transitions occur successively at 112 
and 140 K, releasing nearly the same amount of entropy.  The lower-temperature phase transition 
(denoted as $T_\textrm{N}$) was ascribed to N\'eel ordering by magnetic measurement, but the cause of 
the higher-temperature one (denoted as $T_\textrm{S}$) remained a mystery from the experiment 
\cite{sakurai,sugiyama,nozaki}.  

Then, using the density-functional-theory (DFT) based electronic structure calculations, 
Ishikawa \textit{et al.} \cite{ishikawa} have shown that the crystal field level of nondegenerate $3d_{xy}$ 
orbitals of Cr ion is in fact lower in energy than that of doubly degenerate $3d_{yz}$ and $3d_{xz}$ orbitals, 
giving rise to the orbital degrees of freedom in the system with a $3d^2$ electron configuration.  Thereby, 
they have argued that the higher (lower) temperature phase transition is caused by the ordering of the orbital 
(spin) degrees of freedom of the system.  Because the CrO$_6$ octahedron is elongated along the $c$-axis 
of the crystal structure, this result offers a rare example of the reversal of the crystal-field splitting in 
transition-metal compounds.  

A natural question that arises in this respect would then be what occurs if external pressures are 
applied to this system.  
Recently, Yamauchi \textit{et al.} \cite{yamauchi} reported that the successive phase transitions disappear 
at a high pressure above a few GPa, leaving only one phase transition that divides the system into two 
phases, a high-temperature paramagnetic insulating phase and a low-temperature antiferromagnetic 
insulating phase.  This result can readily be understood if we assume that the reversal of the crystal-field 
splitting ceases to occur under high pressures, so that the lowest doubly degenerate $t_{2g}$ orbitals 
are occupied by two electrons, forming an $S=1$ spin due to Hund's rule coupling, which leads to the 
antiferromagnetic N\'eel ordering of $S=1$ spins at $T_\textrm{N}$, without any orbital ordering.  

In this paper, to check the validity of this assumption, we apply the DFT-based electronic structure 
calculations using the generalized gradient approximation (GGA); in particular, we use the GGA+$U$ 
method for a better description of electron correlations.  We thus show that the reversed crystal-field 
splitting in $\alpha$-Sr$_2$CrO$_4$ is actually restored under high pressures, resulting in the elimination 
of the orbital degrees of freedom of the system.  

\section{Computational details}

We employ the WIEN2k code \cite{wien2k} based on the full-potential linearized augmented-plane-wave 
method for our DFT calculations.  Here, we present results obtained in the GGA for electron 
correlations using the exchange-correlation potential of Ref.~\cite{perdew}.  To improve the description 
of electron correlations in the Cr $3d$ orbitals, we also use the rotationally invariant version of the GGA+$U$ 
method with the double-counting correction in the fully localized limit \cite{anisimov,liechtenstein}.  
In particular, we examine the $U$ dependence of the crystal-field splitting.  The spin polarization is 
not allowed in the present calculations.  The spin-orbit interaction is not taken into account.  
We use the crystal structure measured at high pressures \cite{yamauchi}, which has the tetragonal 
symmetry (space group $I4/mmm$) with one (two) crystallographically inequivalent Cr (O) ions, but we 
apply the local structural relaxations keeping the measured lattice constants unchanged.  We assume 
the high-temperature metallic phase, allowing for no antiferromagnetic spin polarizations.  
In the self-consistent calculations, we use 99 $k$-points in the irreducible part of the Brillouin zone.  
Muffin-tin radii ($R_\textrm{MT}$) of $\sim$2.3 (Sr), $\sim$1.9 (Cr), and $\sim$1.7 (O) Bohr 
depending on pressures are used and a plane-wave cutoff of $K_\textrm{max}=8.00/R_\textrm{MT}$ is assumed. 

\begin{figure}[bht]
\begin{center}
\includegraphics[scale=0.6]{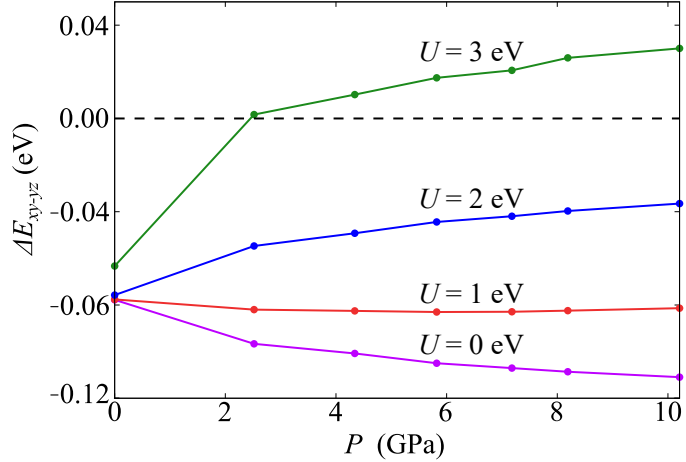}
\caption{Calculated splitting $\Delta E_{xy-yz} = E(d_{xy}) - E(d_{yz})$ of the energy levels of the 
maximally localized Wannier orbitals as a function of applied pressure $P$,  
where several $U$ values are assumed in the GGA+$U$ calculations.}
\label{fig1}
\end{center}
\end{figure}

\section{Results of calculation}

The calculated energy dispersions of the three bands near the Fermi level are fitted by the tight-binding 
bands of the three molecular orbitals obtained as the maximally localized Wannier functions 
\cite{mostofi,kunes}.  The energy-level splittings of the maximally localized Wannier orbitals are thus 
determined as a function of the applied pressure.  

Then, our calculated results for the crystal-field splitting are given in Fig.~\ref{fig1}, where 
we present the pressure dependence of the difference in the energy levels between the 
nondegenerate $3d_{xy}$ orbital and the doubly degenerate $3d_{yz}$ and $3d_{xz}$ 
orbitals of a Cr ion, i.e., $\Delta E_{xy-yz} = E(d_{xy}) - E(d_{yz})$.
Here, we first confirm that the reversal of the crystal-field splitting, 
$E(d_{xy}) < E(d_{yz}) = E(d_{xz})$, actually occurs at ambient pressure, as was found 
in Ref.~\cite{ishikawa}.  Then, under high pressures $P\gtrsim 2$ GPa, we find that the reversal 
of the crystal-field splitting is restored, i.e., $E(d_{xy}) > E(d_{yz}) = E(d_{xz})$, in particular 
when we assume the standard value of $U=3$ eV for Cr ion.  
We should note that the restoration of the reversed crystal-field splitting occurs only for 
larger values of $U$ ($\gtrsim 3$ eV), which means that the restoration is not a simple 
pressure effect but rather the cooperation with $U$ is essential.  Thus, the effect of 
strong electron correlations is suggested to be important in $\alpha$-Sr$_2$CrO$_4$, 
as was noticed in Ref.~\cite{ishikawa}.  More precisely, the repulsive interaction $U$ in 
the GGA+$U$ type of approximations in general works to lower (raise) the energy of 
the occupied (unoccupied) orbitals \cite{liechtenstein,dudarev}.  In the present case, 
the interaction $U$ works to stabilize the state where the doubly degenerate orbitals 
($d_{yz}$ and $d_{xz}$) are occupied by two electrons, in comparison with the state 
where the nondegenerate $d_{xy}$ orbital is occupied by two electrons.  

Thus, we have shown that the reversed crystal-field splitting in $\alpha$-Sr$_2$CrO$_4$ is 
actually restored under high pressures, resulting in the elimination of the orbital degrees 
of freedom of the system, which naturally leads to the single phase transition that divides 
the system into two phases, a high-temperature paramagnetic Mott insulating phase and 
a low-temperature antiferromagnetic insulating phase, in agreement with experiment 
\cite{yamauchi}.  Note that, in the present calculation, the paramagnetic metallic state 
without antiferromagnetic spin polarization is assumed for extracting the values of the 
crystal-field splitting because the paramagnetic Mott insulating state cannot be obtained 
in the GGA+$U$ type of calculations.  However, the present assumption is usually sufficient 
for extracting the values because they are not strongly affected by the spin polarization 
of the system \cite{ishikawa}. 

\section{Summary}

In our previous paper \cite{ishikawa}, we have shown that the successive phase 
transitions observed in a layered perovskite $\alpha$-Sr$_2$CrO$_4$ at ambient 
pressure are ascribed to the active orbital degrees of freedom of the system caused 
by the reversal of the crystal-field splitting of the system.  
However, a recent high-pressure experiment \cite{yamauchi} has shown that the 
successive phase transitions disappear under high pressures, leaving only one phase 
transition dividing the system into two phases, a high-temperature paramagnetic 
insulating phase and a low-temperature antiferromagnetic insulating phase.  

Motivated by this experimental finding, we have made the DFT-based electronic structure 
calculations for $\alpha$-Sr$_2$CrO$_4$ under high pressures in the GGA+$U$ method 
and have demonstrated that the reversal of the crystal-field splitting is actually restored 
under high pressures, so that the orbital degrees of freedom of this system disappears, 
resulting in the single antiferromagnetic phase transition.  

Our result for $\alpha$-Sr$_2$CrO$_4$ on the one hand provides an interesting example 
of the pressure effects in strongly-correlated transition-metal compounds, but on the other 
hand reinforces the idea that the reversal of the crystal-field splitting found in 
$\alpha$-Sr$_2$CrO$_4$ is a rare and fragile phenomenon easily destroyed by the 
external perturbation.  

\bigskip\noindent
\textbf{Acknowledgments}
\par\bigskip\noindent
This work was supported in part by Grants-in-Aid for Scientific Research from JSPS 
(Projects No.~JP17K05521, No.~JP17K05530, No.~JP19K14644, and No.~JP19J10805), 
by a JST Mirai Program (JPMJMI18A3), and by Keio University Academic Development 
Funds for Individual Research.

\end{document}